\documentclass[a4paper,7pt,twoside]{CoAst}
\usepackage{epsf,graphicx,natbib,fancyhdr}

\renewcommand{\chaptermark}[1]{\markboth{#1}{}}

\newcommand{\kopf}{\small\sf\itshape Comm. in Asteroseismology \\ Vol. number, publication date (will be inserted in the production process)}

\newcommand{\Authors}[1]{\begin{center}\normalsize\bf\sf #1 \end{center}}

\renewcommand{\author}[1]{\begin{center}\normalsize\bf\sf #1 \end{center}}
\newcommand{\Address}[1]{\begin{center}\small\sf #1 \end{center}}

\newcommand{\Accepted}[1]{{\vspace{2mm}\small \noindent  \hspace*{0mm}Accepted: } \small #1 \normalsize}

\renewenvironment{abstract}{\section*{Abstract}\normalsize\sf}{}
\newcommand{\References}[1]{\begin{flushleft}{\large References\\}\vspace*{2mm}\small #1 \end{flushleft}}

\newcommand{\chapterCoAst}[3]{\chapter[\sf\normalsize #1\\ \footnotesize \hspace*{5mm}by #2 \sf\normalsize][]{#1\\}\rhead[\fancyplain{}{\sf\footnotesize \center{#3}}]{\fancyplain{}{\sffamily\thepage}}\lhead[\fancyplain{\kopf}{\sffamily\thepage}]{\fancyplain{\kopf}{\sf\footnotesize \center{#2}}}}

\newcommand{\figureDSSN}[5]{\begin{figure}[#4]
\centering
\includegraphics*[#5]{#1}
\caption{#2}
\label{#3}
\end{figure}}

\renewcommand{\chaptername}{}

\newcommand{\acknowledgments}[1]{\vspace*{5mm}\noindent  \textbf{Acknowledgments.} #1}

\usepackage[margin=3cm,includehead,includefoot,heightrounded,a4paper]{geometry}

\usepackage{amssymb,amsmath} 

\usepackage{url}

\include{journals}
\bibpunct[; ]{(}{)}{;}{a}{}{}

\def\rfr{\smallskip\par\noindent
        \hangindent=7truemm
        \hangafter=1}

\newcommand{\echelle}{\'{e}chelle~}
\newcommand{\Echelle}{\'{E}chelle~}

\begin{document}
\sf
\chapterCoAst{A comparison of Bayesian and Fourier methods for frequency determination in 
asteroseismology}
{T.\,R.\,White, B.\,J.\,Brewer, T.\,R.\,Bedding, et al.}{A comparison of Bayesian and Fourier methods 
for frequency determination}
\Authors{T.\,R.\,White$^1$, B.\,J.\,Brewer$^{1,2}$, T.\,R.\,Bedding$^1$, D.\,Stello$^1$ and H.\,Kjeldsen$^3$} 
\Address{$^1$ Sydney Institute for Astronomy (SIfA), School of Physics, University of Sydney, NSW 2006, Australia\\
$^2$ Department of Physics, University of California, Santa Barbara, CA 93106-9530, USA\\
$^3$ Danish AsteroSeismology Centre (DASC), Department of Physics and Astronomy, Aarhus University, DK-8000 Aarhus C, Denmark
}

\noindent
\begin{abstract}
Bayesian methods are becoming more widely used in asteroseismic analysis. In particular, they are being used
to determine oscillation frequencies, which are also commonly found by Fourier analysis. It is
important to establish whether the Bayesian methods provide an improvement on Fourier methods. We compare, using
simulated data, the standard iterative sine-wave fitting method against a Markov Chain Monte Carlo (MCMC) code that has
been introduced to infer purely the frequencies of oscillation modes (Brewer et al. 2007). A uniform prior probability 
distribution function is used for the MCMC method. We find the methods do equally well at determining the correct 
oscillation frequencies, although the Bayesian method is able to highlight the possibility of a misidentification due
to aliasing, which can be useful. In general, we suggest that the least computationally intensive method is preferable. 
\end{abstract}

\Accepted{June 7, 2010}

\section*{1.   Introduction}
Bayesian methods are increasingly being used for asteroseismic analysis. Most effort has been directed at extracting
mode parameters by fitting to the Fourier power spectrum (e.g Appourchaux 2008; Benomar 2008; Benomar et al. 2009;
Gaulme et al. 2009), but there have also been applications that
involved fitting directly to the time series (Brewer et al. 2007; Brewer \& Stello 2009). It is the latter approach,
which can be thought of as an alternative to calculating the power spectrum, that is the subject of this paper.
Determining the frequencies at which stars oscillate is fundamental to asteroseismology. The first step in
doing so is generally to calculate the power spectrum of the time series, in which the frequencies of 
oscillation will appear as peaks. However, this is complicated by noise and
aliasing so that it is not always immediately obvious which peaks are real. In this paper, we use simulated data to 
determine if a Bayesian Markov Chain Monte Carlo code is more effective at determining real frequencies than
a standard iterative sine-wave fitting code.

\subsection*{1.1.  Iterative sine-wave fitting}
To systematically extract the peaks that are most likely to be real, an iterative algorithm (Roberts et al. 1987) is 
commonly used (e.g. Carrier \& Bourban 2003; Kjeldsen et al. 2005; Bedding et al. 2007). This is referred to as iterative 
sine-wave fitting (also called CLEAN). In this algorithm, the discrete Fourier transform of the time series is 
calculated, generating the amplitude spectrum. The highest peak in the amplitude spectrum is identified and the 
corresponding sinusoid is then subtracted from the time series. The discrete Fourier transform of the residual time 
series is then calculated, and the process is repeated until the amplitude of the highest peak is below a chosen 
threshold. The result is a list of frequencies, amplitudes and phases that account for most of the variations in the 
time series. 

However, this technique may run into difficulties. It is possible for alias peaks (sidelobes) to be enhanced by noise to
the point where they have a higher amplitude in the Fourier spectrum than the genuine peak. In this case, iterative 
sine-wave fitting programs will identify the alias and subtract it from the time series. The amplitude of the genuine
peak in the Fourier spectrum will be diminished by this process to the point where it may not be found at all.
This problem may be exacerbated if the separation between oscillation peaks is approximately an integer multiple of the 
typical period between gaps in the data (usually one cycle per day in ground-based data) since the alias peaks may 
reinforce. In addition to this, noise peaks may obscure the signal. 

Other concerns extend from the nature of Fourier methods. Bretthorst (1988) has shown that the power spectrum is an
optimal procedure only in the case where only one mode is present, or if multiple, they are well separated in frequency. 
In practice, particularly with solar-like oscillations,
this is rarely the case, with multiple modes often closely spaced in frequency. A further consideration is that the
oscillation modes will vary in time such that a single mode must be represented by multiple sinusoids in the Fourier
series. An additional problem can arise from the nature of the algorithm itself: when a frequency is found, and its
amplitude and phase fitted to the time series, the corresponding sinusoid is subtracted. However as these parameters
cannot be exact, biases are introduced into the residual time series.

\subsection*{1.2.  Bayesian Methods}

The above concerns have led to the consideration of Bayesian methods for determining frequencies.
Probability theory may be used as a mathematical model of our belief in the plausibility of various hypotheses. Our
knowledge of a set of parameters, \(\theta\), given prior information and assumptions, \(I\), is represented by the \em
prior probability distribution\em, \(p(\theta|I)\). Given new data, \(D\), \em Bayes' theorem\em,
\begin{equation}
p(\theta|D,I)\propto p(\theta|I)p(D|\theta,I), \label{Bayes}
\end{equation}
tells us about our new state of knowledge, the \em posterior distribution\em. The distribution \(p(D|\theta,I)\), the
probability distribution for the data given the parameters, is a measure of how well the data are predicted by the model.
In the case where the data are known and fixed, \(p(D|\theta,I)\) becomes dependent on \(\theta\) only and is called the
\em likelihood function\em.  How much we believe our model depends on both how well we originally believed it (the prior
distribution), and on how well it predicts the new data. It is important to note that probabilities are always
conditional on the underlying background information and/or assumptions \(I\), even when these are not explicitly
stated. Bayes' theorem provides a means of finding the most probable model that could produce the observed data.

A method using this probabilistic reasoning has been developed by Brewer et al. (2007) and applied to the subgiant stars
\(\nu\) Indi (Bedding et al. 2006; Carrier et al. 2007) and \(\beta\) Hydri (Bedding et al. 2007). This method utilises 
a version of the Metropolis-Hastings algorithm (Neal 1993), itself a Markov Chain Monte Carlo (MCMC) algorithm (Gregory 
2005), to determine the most likely set of frequencies (represented by \(\theta\) above) that could give rise to
observed data. Essentially, the code samples the \(\theta\) parameter space -- the number of frequencies and their
values -- through a random walk with more time being spent in regions of higher probability. At each iteration of the
code, the current state is randomly perturbed. If the perturbation results in a higher posterior probability then it is
accepted as the new state, otherwise the perturbed state is accepted with a probability proportional to the ratio of the
new posterior probability density to the old.

The method of Brewer et al. (2007) analyses the observed time series directly. To do this it finds the most likely
sinusoids that fit the time series, assuming that the time series is composed of a small number of sinusoidal signals 
(one per mode) and 
Gaussian noise. This assumption is not generally valid due to the stochastic nature of excitation and damping of the
oscillations. Some methods that take this in to account are able to infer frequencies and line widths by fitting to the
power spectrum (eg. Appourchaux 2008). However, the presence of gaps in the data, and the stochastic nature of
oscillations results in the possibility that information will be lost in the power spectrum (Bretthorst 1988). 
It was this potential
loss of information from using the power spectrum that was one of the reasons for considering Bayesian methods in the 
first place. The convenience of the power spectrum cannot be doubted and for data with good coverage and long mode 
lifetimes, the loss of information is negligible. Nevertheless it would be ideal to have a method that both analyses the
time series directly and takes into account that the oscillations are not purely sinusoidal. Such a method has been
developed (Brewer \& Stello 2009), but is unfortunately computationally intensive and is currently only practical for short time
series (fewer than $\sim1500$ points). For this reason, it is the method of Brewer et al (2007) we use in our comparison here with
Fourier methods.

A major (although potentially dangerous) feature of Bayesian inference is the ability to incorporate extra knowledge in
determining which parameters are more likely by choosing a descriptive prior distribution. For some stellar
oscillations, namely high-overtone, low-degree acoustic oscillations in spherically symmetric stars, modes are usually
expected to follow the asymptotic relation,
\begin{equation}
\nu_{n,l} = \Delta \nu (n + \frac{1}{2} l + \epsilon) - l(l+1)D_0, \label{asympf}
\end{equation}
where \(\Delta\nu\) is called the large separation and depends on the sound travel time across the whole star, \(D_0\)
is sensitive to the sound speed near the core and \(\epsilon\) is sensitive to the surface layers (Tassoul 1980, 1990). With this in mind, a prior distribution could be chosen in which a regular spacing of modes is
anticipated. The MCMC code would then favour finding peaks with a regular spacing over finding peaks that do not fit
this pattern. This was a feature of the code when it was introduced by Brewer et al. (2007) and tested on simulated data 
(which did have a constant separation between modes) and when it was used on real observations of \(\nu\) Indi by
Bedding et al. (2006) to find the large separation. The value inferred by the Bayesian method for the large
separation, \(\Delta\nu=24.25\pm0.25\mu\)Hz, agreed well with the value obtained from the peak of the autocorrelation
function of the power spectrum, a Fourier method. 

As mentioned, this use of a descriptive prior distribution can be dangerous. Although the modes of oscillation will be
separated in frequency by approximately equal amounts, in general there will be a departure from the asymptotic
relation. The large separation itself may be frequency-dependent. By using a prior that is too prescriptive, this may be
missed by the Bayesian method. To avoid this possibility, the method described by Brewer et al. (2007) incorporated a
uniform component and a regular pattern of Gaussian peaks, as opposed to delta functions. However, this may still be too
prescriptive. A further analysis of the data on \(\nu\) Indi showed that the large separation was indeed a function of
frequency, with the average large separation of \(\Delta\nu=25.14\pm0.14\mu\)Hz larger than first realised 
(Carrier et al. 2007).

Further use of this method was made by Bedding et al. (2007) on \(\beta\) Hydri. They used the Bayesian method as a supplement
to traditional Fourier methods to determine individual frequencies, although no assumptions about the frequency
distribution were made. To avoid detecting noise peaks in the Fourier analysis, only the peaks with a signal-to-noise
ratio (S/N) \(\ge\) 4 were included. Comparing the results with the Bayesian method, one extra peak was found by the
Bayesian method, lying precisely on the \(l=1\) ridge of the \echelle diagram. The same peak was found by the Fourier
iterative sine-wave fitting with S/N = 3.0.

These studies of \(\nu\) Indi and \(\beta\) Hydri have obtained similar results using both Bayesian and traditional 
Fourier method. The question arises as to which method, if either, is superior. Brewer et al. (2007) suggested that the Bayesian
method was less susceptible to aliasing and, by effectively fitting multiple sinusoids simultaneously, was less
susceptible to any issues caused by the subtraction of modes from the time series. On the other hand, the Bayesian
method is more computationally intensive for large data sets, and it has been suggested that its advantage over the
traditional Fourier methods is with shorter, noisier, and more incomplete data sets. Despite these presumed advantages,
a detailed comparison of the methods has not yet been made. It is the purpose of this paper to do this comparison on
simulated data for which the frequencies are known.

\section*{2.   Programs being tested}

Two different iterative sine-wave fitting programs were used, both written by one of us (HK). They differ in that 
one program re-adjusts the frequencies, amplitudes and phases of all previously extracted
peaks at each iteration, in an attempt to improve the fit. However, there is a time penalty to this, and it may not
necessarily lead to improved results because the process may be unstable. Hereafter, the version which does not 
recalculate parameters at each iteration is referred to as \em fast \em and the other as \em slow\em.

The program used to evaluate Bayesian methods of determining frequencies was written by one of us (BJB) and is a
variation of the program introduced by Brewer et al. (2007), which implements a Markov Chain Monte Carlo (MCMC) algorithm.
The principles of the original program have been discussed above. Here, we outline how our implementation of the code
differs.

The most important difference is the choice of the prior probability distribution. Our choice here can have a
significant impact on the results obtained. Brewer et al. (2007) used a prior distribution which took into account that the
frequencies of low-degree $p-$mode oscillations are expected to be approximately given by Equation \ref{asympf}. This is
not a valid assumption if a mode exhibits mixed $p-$ and $g-$mode behaviour, as in evolved stars, or if the large
separation, \(\Delta \nu\), is not constant with frequency. Setting this prior will have a significant impact on the 
posterior distribution, potentially placing unreasonably
high confidence in low signal-to-noise peaks that fit a regular spacing, and down-weighting real peaks that do not. For
our implementation of this method, we have chosen only to use a uniform prior that does not anticipate the separation of
peaks to avoid the possibility of the code finding a regular spacing that may not be there. The uniform prior is given by
\begin{equation}
g = \frac{1}{\nu_{\mathrm{max}}-\nu_{\mathrm{min}}}, \label{prior}
\end{equation}
where \((\nu_{\mathrm{min}},\nu_{\mathrm{max}})\) is the range of frequencies over which we attempt to find oscillations.

Since Brewer et al. (2007) used a sharply peaked prior, it was possible for the program to become stuck in a local minimum and
not adequately sample the parameter space. To overcome this problem parallel tempering was used with respect to the prior.
In our case we do not use parallel tempering since the uniform prior is without the sharp peaks that could cause the chains 
to become stuck in local minima.

Will this change to the prior have much impact on the number of frequencies used in the models at each iteration? That is,
how effective was the use of a prior that expected a regular spacing of peaks at suppressing peaks that did not meet this
criteria? Figure 3 of Brewer et al. (2007) shows the output of a MCMC code run on simulated data that contained 17 
frequencies. The code infers that there are at least 16 frequencies, and possibly up to about 20, with 16 the most probable
solution. A typical output of the MCMC code used in this paper is shown in Figure \ref{figure1} for the simulated data
discussed in Section 3. After a short initial burn-in period, the distribution of the number of frequencies
settles into the posterior distribution. The number of frequencies fluctuates between approximately 50 and 150, averaging
around 90. The actual number used was 61. Although our simulated data is different to that of Brewer et al. (2007), it does 
appear that the uniform prior results in additional frequencies being accepted than with the descriptive, regularly peaked
prior, as could be expected. Since we do expect there to be significant departures from an equal spacing in some stars, 
we are prepared to accept the increased chance of detecting noise peaks and aliases between equally spaced frequencies for 
this comparison with Fourier methods.

\figureDSSN{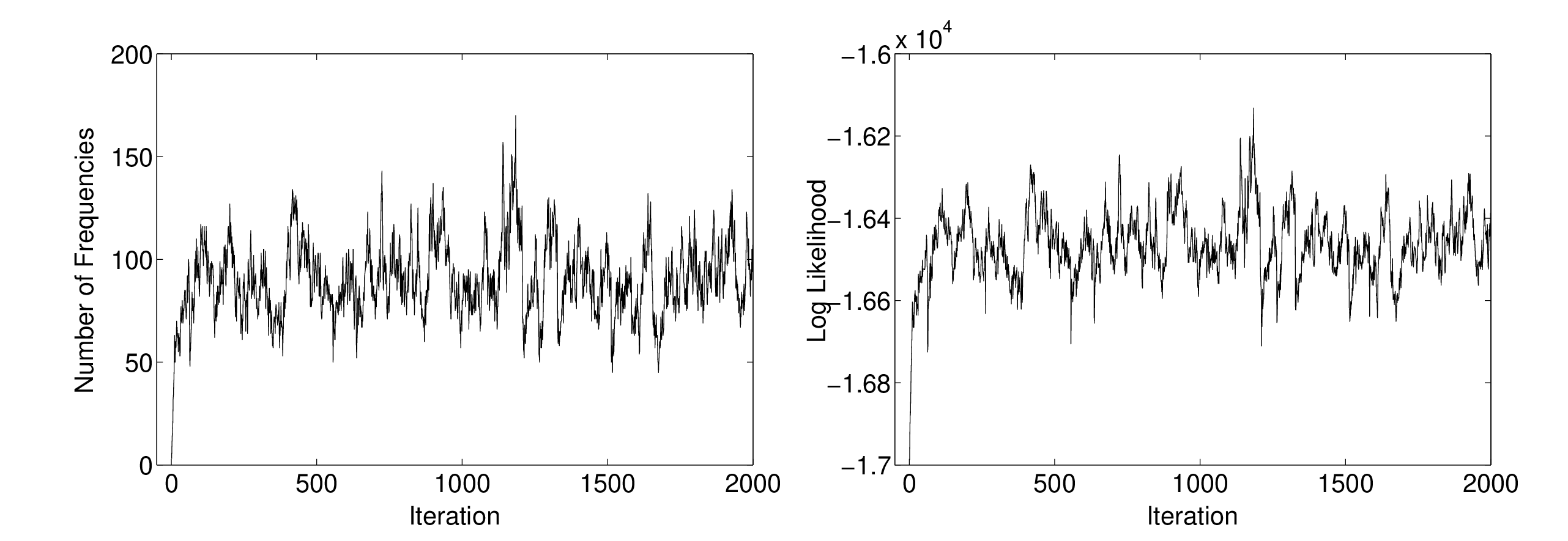}{Results of a MCMC run with a uniform prior. \em Left \em Number of frequencies fitted as a function 
of iteration. \em Right \em Log likelihood of the proposed model as a function of iteration.}
{figure1}{t}{clip,angle=0,width=150mm}

\pagebreak
The original code, in comparison to traditional methods, had the deficiency of not returning the amplitudes of the
modes, but only their relative probabilities. This has since been rectified, with the Markov Chain now sampling the
distribution of both frequencies and amplitudes at each iteration.

\section*{3.  Tests with coherent oscillations}\label{coherent}

Our first test of the algorithms was the simplest case possible, in which the oscillations are pure sinusoids and
are well-separated in frequency. While no stellar oscillation is entirely coherent, there exist many cases where the
mode lifetime is long enough for this to be a good approximation. For this test it was important to probe a variety of
signal-to-noise (S/N) ratios to gauge the performances of the programs at different levels. S/N ratios from 2 and 5 
were probed in this test.

\figureDSSN{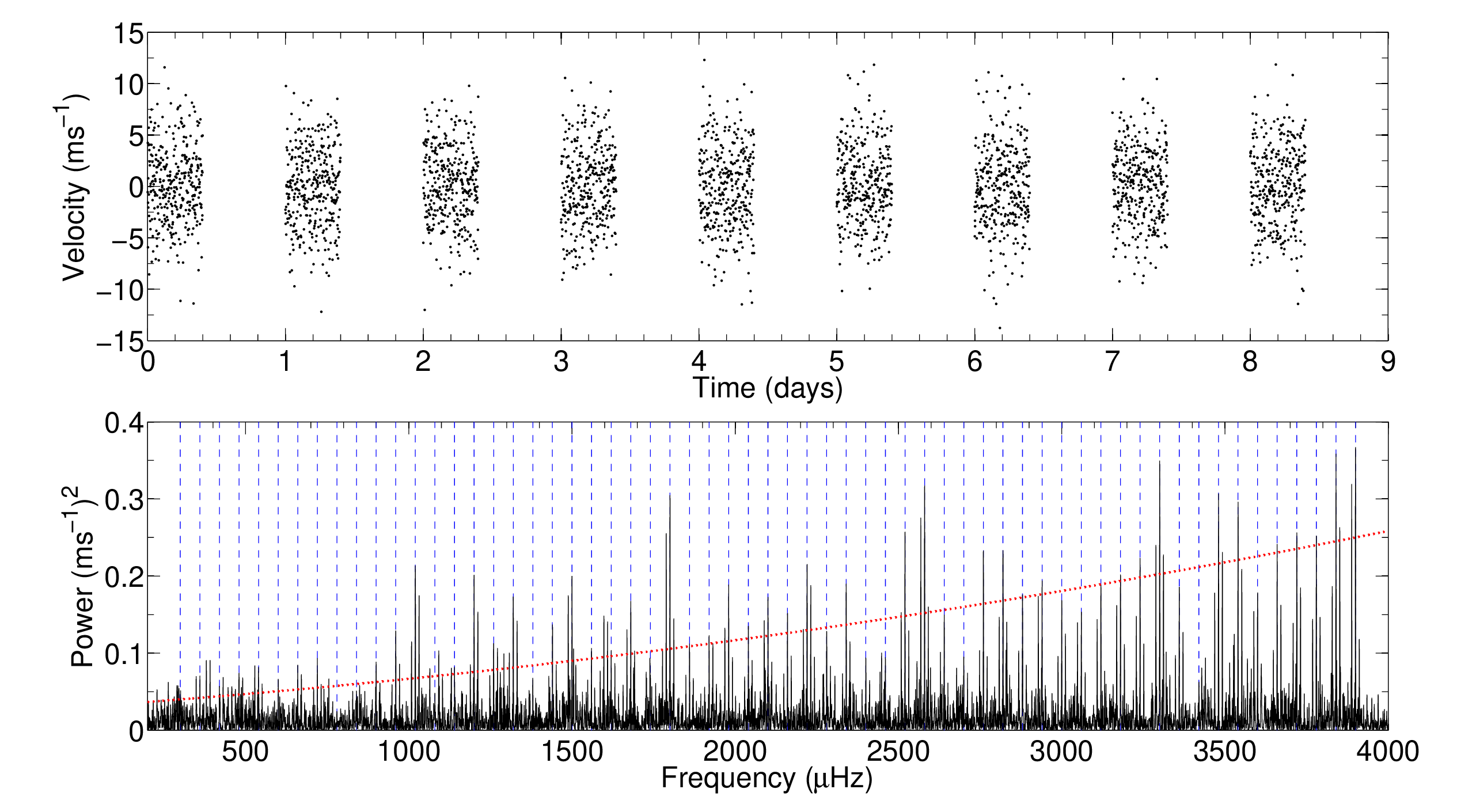}{\em Top \em Simulated time series with coherent oscillations ranging in S/N from 2 to 5. \em Bottom
\em Corresponding power spectrum. Input frequencies are indicated by the vertical blue dashed lines. Input power 
as a function 
of frequency is indicated by the red dotted line. Sidelobes at \(\pm11.6\mu\mathrm{Hz}\) due to daily gaps are 
clearly visible. Colour available on online version.}{figure2}{t}{clip,angle=0,width=150mm}

The time series was generated as a sum of 61 sinusoids with frequencies separated by 60 \(\mu\)Hz, ranging from 300 
\(\mu\)Hz to 3900 \(\mu\)Hz. Phases were chosen at random. The generated signal was sampled every 100 seconds for a 
total of 9 days. Gaps are common in astronomical datasets, primarily due to the usual restriction of observations 
to be taken at
night, which causes the aliasing that is so problematic to frequency analysis. To simulate this, only the first 40 per
cent of a `day' of simulated observations was retained as part of the time series. The resulting time series had 3114
data points. Random Gaussian noise with a standard deviation of 3.09 ms\(^{-1}\) was added to the data, which was
calculated to provide an average noise in the amplitude spectrum of 0.10 ms\(^{-1}\) 
(from equations A1 and A2 of Kjeldsen \& Bedding 1995). The amplitudes chosen for the sinusoids varied linearly from 
0.2 ms\(^{-1}\) at 300 \(\mu\)Hz up to 0.5 m
\(^{-1}\) at 3900 \(\mu\)Hz, to give the desired range of S/N ratios. It should be noted that, due to the effects of
noise, the measured S/N of these peaks will differ from the input values, with some peaks being suppressed and
others enhanced. The final time series is shown in Figure \ref{figure2}, together with its power spectrum. It is
apparent that many more peaks are present in the power spectrum than the 61 input sinusoids marked by dashed lines, with
some due to aliasing occurring at \(\pm11.6 \mu\)Hz (\(\pm1\) cycle/day) from the input peaks and others due to
noise.

\subsection*{3.1.  Results}
The simulated time series was analysed with both versions of the iterative sine-wave fitting program, and with the 
Bayesian MCMC code. The peaks extracted by the fast iterative sine-wave fitting program and the probability spectrum 
output by the MCMC code are shown in Figure \ref{figure3}. It is immediately
apparent from these graphs that both methods successfully identified the modes with the highest S/N and that they agree 
in general. Although the relative heights of peaks are different between the two spectra in 
Figure \ref{figure3}, 
the modes with the highest S/N generally correspond to those with the greatest posterior probability.

\figureDSSN{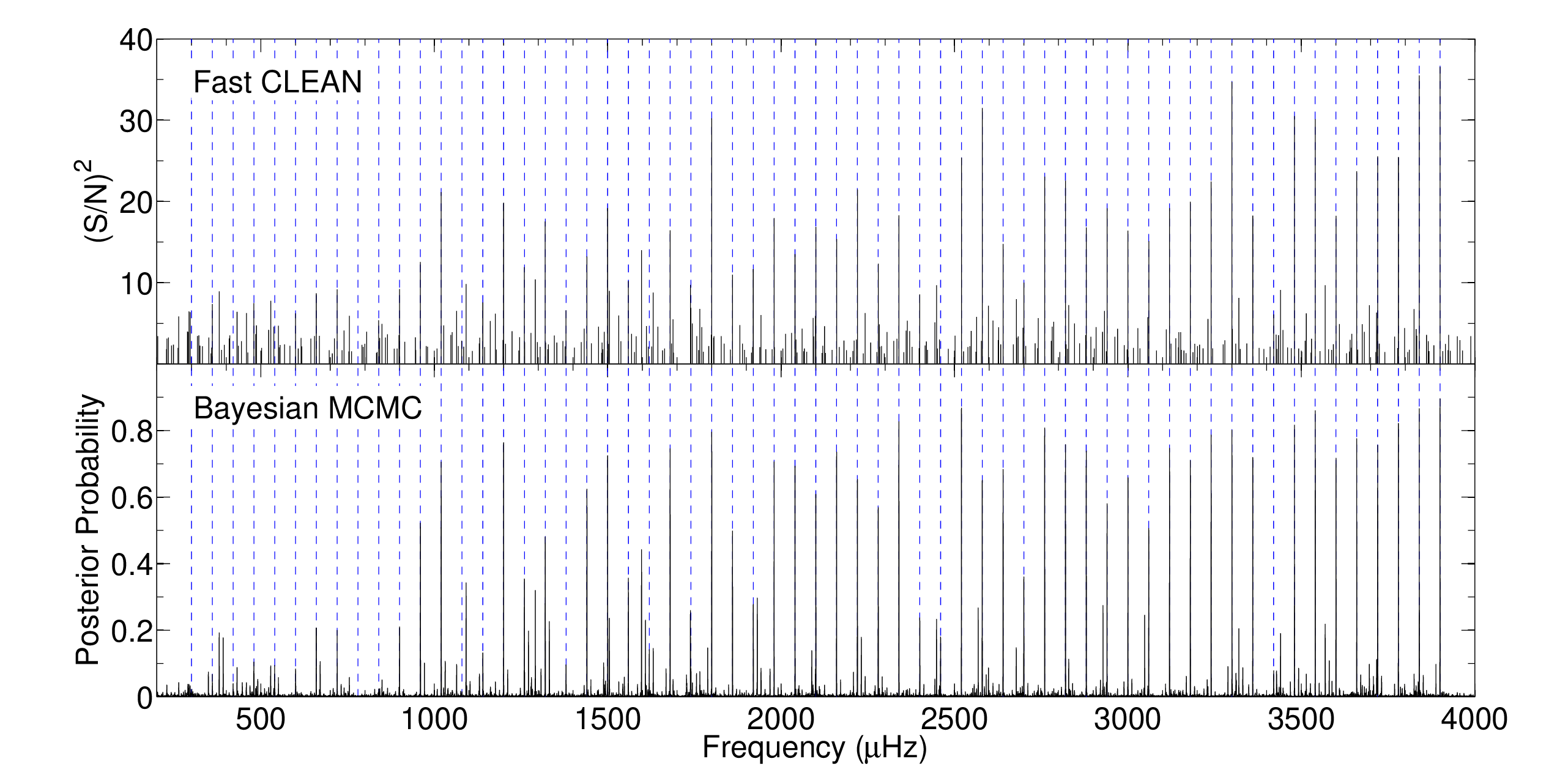}{\em Top \em Peaks extracted by the fast iterative sine-wave fitting program.
\em Bottom \em Probability spectrum from the Bayesian MCMC program. Blue dashed lines indicate the input frequencies.}
{figure3}{t}{clip,angle=0,width=150mm}

To further see this relation between probability and S/N,
Figure \ref{figure4} compares the probability and S/N of matched peaks between the Bayesian and Fourier methods.
Between a S/N of 2 and 4 there is a roughly linear relationship between S/N and posterior probability. Either side of
this region the probability saturates, resulting in a flattening of the graph. From this figure it is apparent that
peaks can generally be believed (open circles) if the S/N is above $\sim$3.5 or the posterior probability is above 
$\sim$0.4. 

If a peak was identified by one program only, whether real, alias or noise, it appears on this graph on the vertical
line of points at a S/N of $\sim$1.1 or the horizontal line of points at a posterior probability of $\sim$0.01. If the
iterative sine-wave fitting program did not find a peak above the S/N$\sim$1.1 threshold it is either due to the peak
having an inherently low S/N, or to the peak being diminished when a sine-wave was removed from the time series in 
previous iterations. This
second explanation is responsible for the number of alias peaks in this category, which would be effectively removed in
the Fourier case if the correct peak were identified first. However, if it happened that the fast Fourier program
mistakenly identified the alias of a given peak, then it would not later detect the correct peak. The slow
Fourier program would have a chance to change its identification, though in practice, this turned out to be rare. 
We found that the Bayesian program is more likely
to identify both, and so may be of use as an alert to the possibility that a peak has been misidentified with 
its alias. Comparing the \echelle diagrams of the fast iterative sine-wave fitting program with the MCMC Bayesian code in Figure \ref{figure5}, we see that they effectively do equally well at identifying the correct peaks.

\figureDSSN{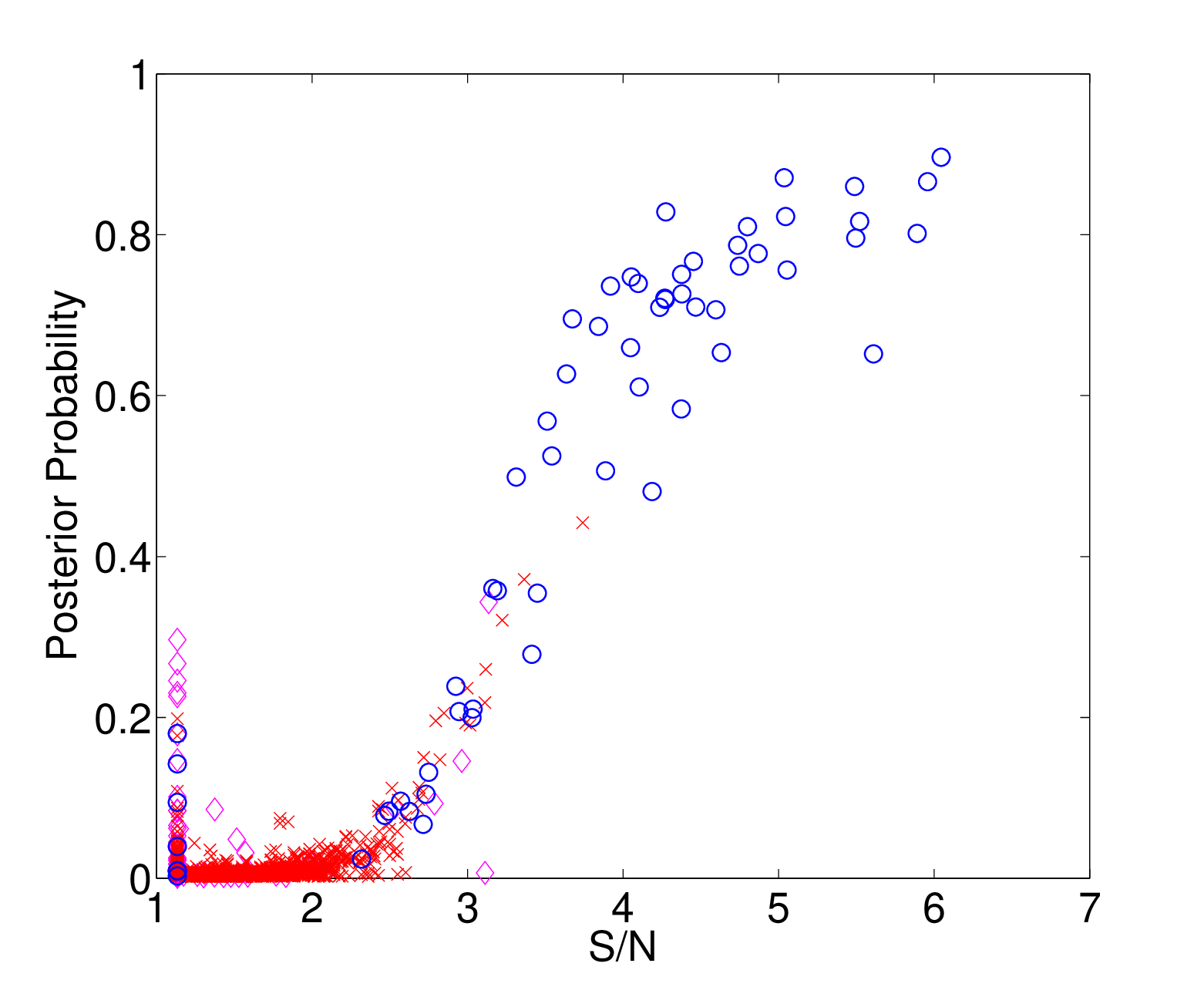}{Posterior probability from Bayesian MCMC versus S/N from fast iterative sine-wave
fitting program. Blue circles correspond to true input frequencies, magenta diamonds are aliases at \(\pm\)11.6\(\,\mu\)Hz, 
and red crosses are noise peaks.}{figure4}{}{clip,angle=0,width=80mm}

\figureDSSN{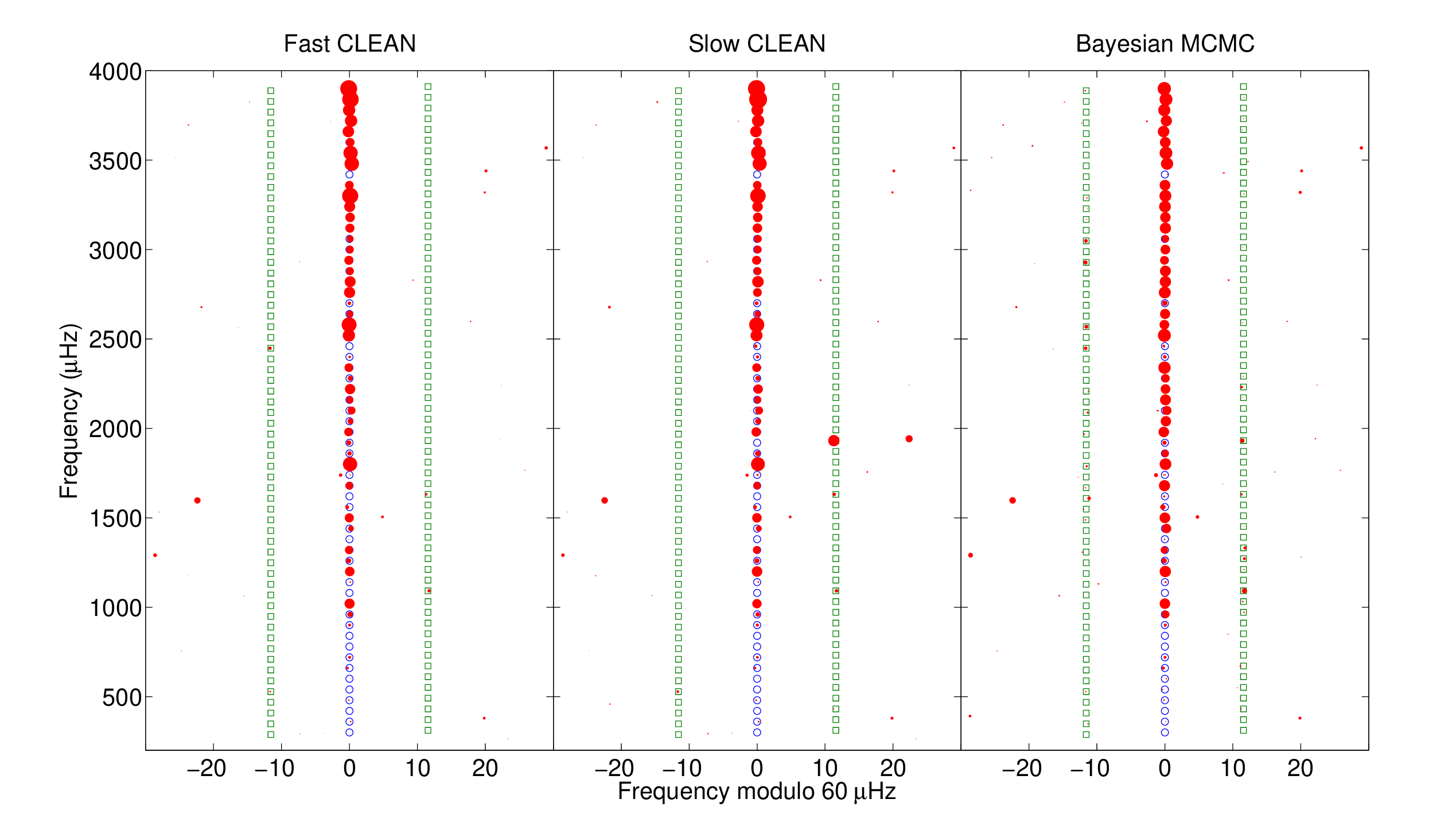}{\em Left \em \Echelle diagram from fast iterative sine-wave fitting program. \em Centre \em
\Echelle diagram from slow iterative sine-wave fitting program. \em Right \em \Echelle diagram from Bayesian MCMC. The
blue open circles indicate the location of real input peaks, with the green open squares indicating their aliases. Red 
filled circles indicate the peaks found by each program, with the size indicating the relative strength of each peak. 
The sizes have been scaled, so that peaks that follow the linear trend in Figure \ref{figure4} will have 
approximately the same size in each diagram.}{figure5}{}{clip,angle=0,width=150mm}

Despite the much longer time required to run the slow version of the iterative sine-wave fitting program, there did not
appear to be any significant difference to the fast version, as can be seen in Figure \ref{figure5}. In general, the
slow version is slightly more accurate in determining the frequencies, although in one case, the slow version has
wrongly chosen the first and second aliases of the correct peak when the fast version did not. Stello et al. (2006)
similarly found that the difference between the two was minor. They found that the fast version identified 1\(\%\) false
peaks compared to a perfect result by the slow version for simulated data from 100 time series with 17 equally-spaced
frequencies with no noise, extracting 10 frequencies per time series. The scatter of the
output frequencies relative to the input was roughly equal to the frequency resolution for the slow version, and about
twice that size for the fast version. For non-coherent oscillations, the differences were diminished. The slow version
takes a factor of \(1.5(N+1)\) longer than the fast version, where \(N\) is the number of
frequencies to be extracted (Stello et al. 2006) and so, for a large number of frequencies, is prohibitively slow.

\section*{4.  Tests with problematic aliasing}

The next test investigates situations where aliasing is particularly problematic, that is, when the frequency separation
between modes is close to an integer multiple of one cycle/day. The first side-lobes of adjacent peaks will coincide 
when the separation is exactly 2 cycles/day. A chirp was introduced to the input frequencies so that the frequency 
separation increased from 1.8 
cycles/day up to 2.2 cycles/day, at a rate of 0.1 cycles/day/order. The input S/N of each peak was fixed to 3.5. 
As previously mentioned, the effects of noise will modulate this in the output time series.

The window function and noise was the same as in the previous test (Section 3). The power spectrum of this time series 
is shown in Figure \ref{figure6}. The aliases are clearly apparent, approximately halfway between the real input 
peaks (dashed lines). Many of the sidelobes are higher than real peaks.

\figureDSSN{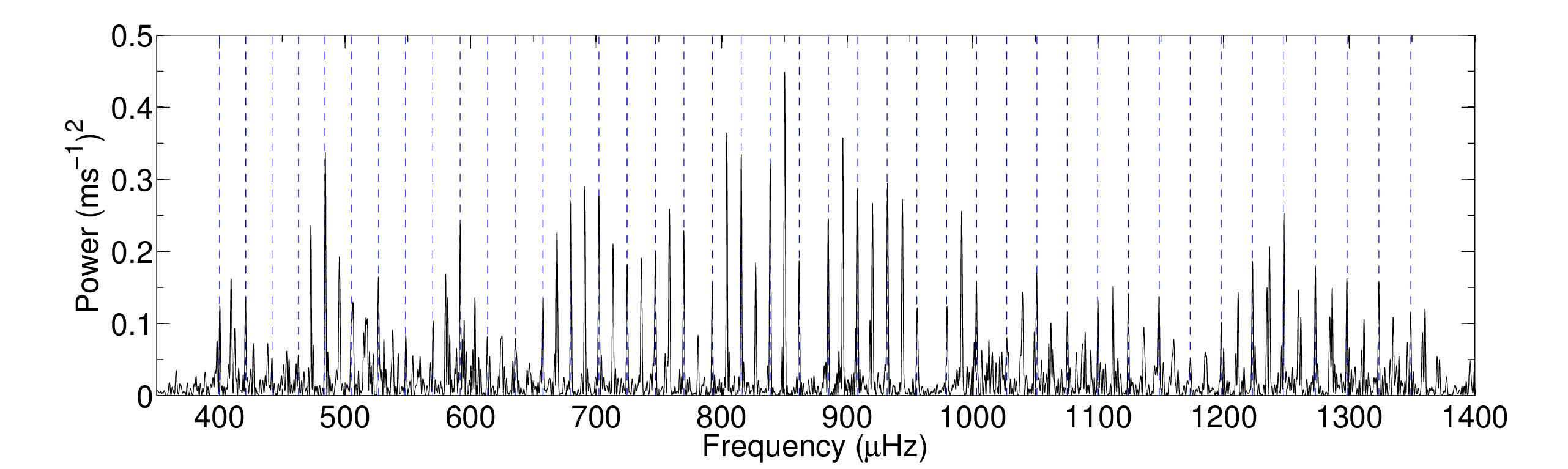}{Power spectrum of simulated time series with problematic aliasing. Input 
frequencies are indicated by the blue dashed lines.}{figure6}{}{clip,angle=0,width=150mm}

\subsection*{4.1.  Results}

Both programs performed quite well when the separation was significantly different from two cycles/day. 
When the aliases of 
adjacent modes coincide (in the middle of Figure \ref{figure6}), both failed to detect the correct frequencies. 
Matching the frequencies extracted by each program and plotting posterior probabilities against S/N in Figure \ref
{figure7}, we see the same general trend as shown in Figure \ref{figure4}, although there is clearly more 
scatter. This scatter is due to the Bayesian MCMC code alternatively sampling both the real and alias peaks, and down-
weighting their relative probabilities, whereas the iterative sine-wave fitting code will identify the highest peak at 
its S/N, real or alias, and miss the other altogether. This explanation is further borne out by the \echelle diagrams, 
shown in Figure \ref{figure8}. The Bayesian method does have an advantage in drawing attention to the possibility 
of confusion, but is not any better at determining which is real.

\figureDSSN{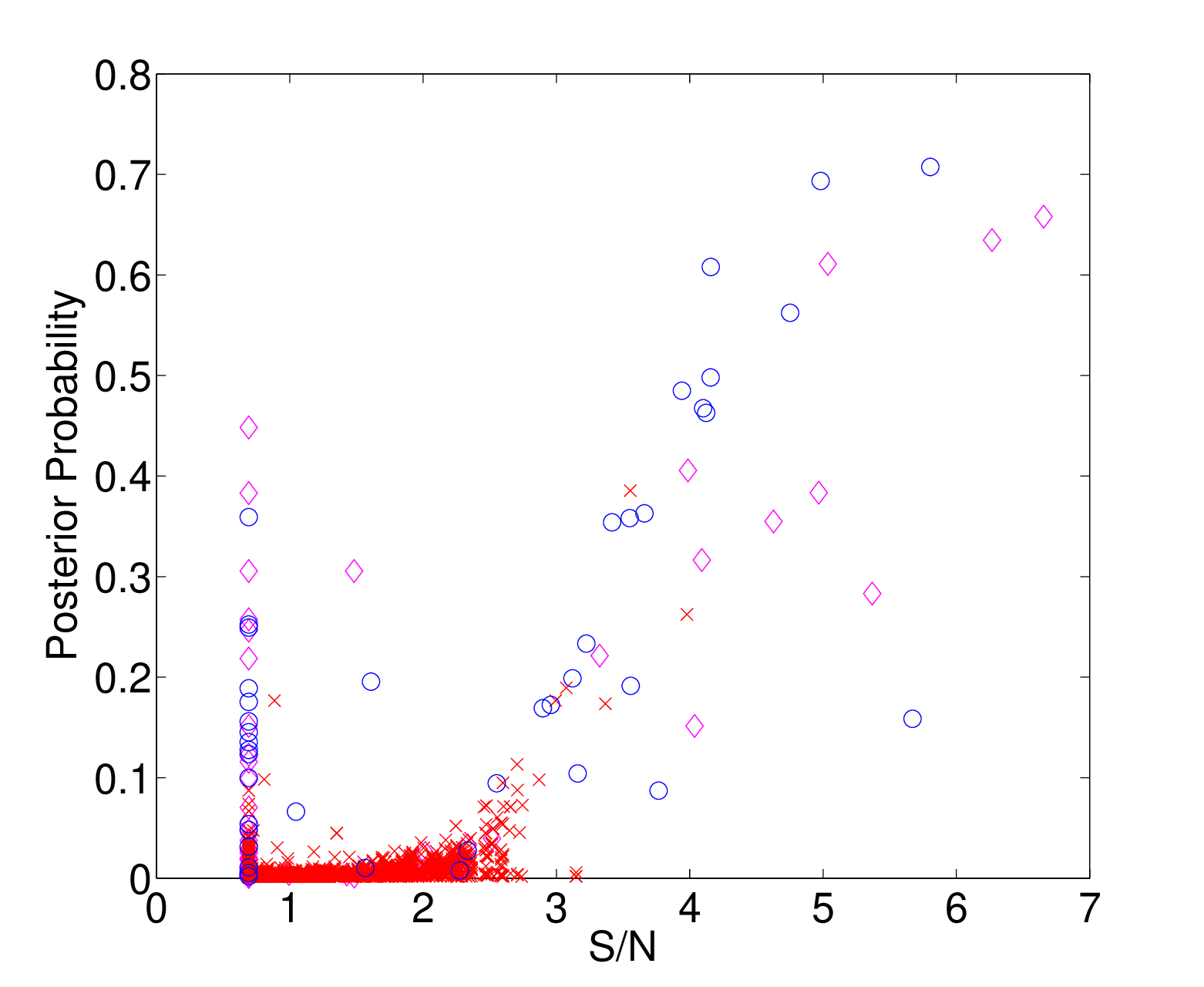}{Same as for Figure \ref{figure4} but for a time series with problematic 
aliasing.}{figure7}{}{clip,angle=0,width=80mm}

\figureDSSN{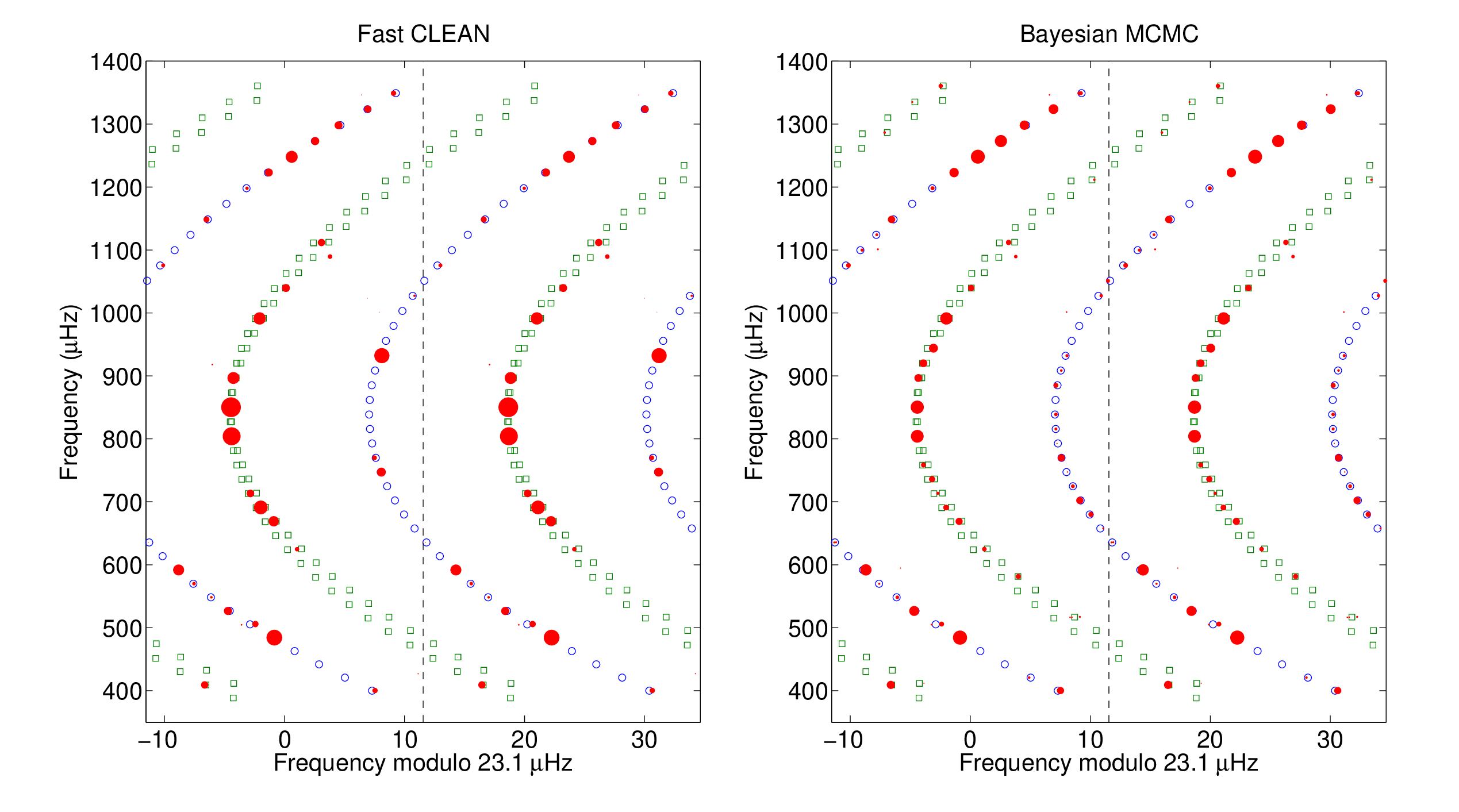}{\Echelle diagrams for a time series with problematic aliasing. \em Left \em Fast iterative 
sine-wave fitting, \em right \em Bayesian MCMC. Symbols same as for Figure \ref{figure5}. Note that the \echelle
diagrams have been plotted twice so that a ridge of input frequencies is continuous, with the dashed line
indicating the half-way join. The ridge of input frequencies, and their aliases are curved due to the varying 
separation of the input series.}{figure8}{t}{clip,angle=0,width=150mm}

\section*{5.   Discussion and Conclusions}
The Bayesian program and the traditional Fourier methods do equally well at identifying the correct frequency of 
stellar oscillations. The Bayesian method is effectively 
attempting to fit multiple sinusoids to the time series, whereas the iterative sine-wave fitting program finds the 
highest peaks, one at a time, from the Fourier amplitude spectrum. This often leads the Bayesian program to sample both 
the real peak and its aliases, whereas the Fourier program will identify only the highest of these. Without a more 
descriptive prior probability, it is not possible for the Bayesian program to avoid identifying alias or 
noise peaks that have high S/N. However, a more detailed prior might cause the results to be dominated by what was 
expected and not by what is actually present. It may be the case that the frequencies that are unexpected prove to be 
the most interesting.

When aliases are strong, the Bayesian method does have an advantage in highlighting the possibility of the confusion 
because it should detect both aliases and real peaks. In general however, we have found no advantage of the Bayesian 
method over the traditional Fourier methods. It is therefore recommended that the least computationally intensive 
program be used. The fast Fourier program was found to be fastest, provided there were not many peaks to be extracted, 
and there are a large number data points (\(>\sim10000\)). If a large number of peaks is required, and there are not too 
many data points, then the Bayesian method takes a comparable time. The slow Fourier program is prohibitively slow for 
determining any more than a few frequencies.

While we have shown that there is no major advantage in using the Bayesian approach discussed here,
it is clear that Bayesian methods will continue to be used for fitting to the power spectrum once it has been calculated 
using traditional Fourier methods.

\acknowledgments{}
We acknowledge support from the Australian Research Council. TRW is supported by an Australian Postgraduate Award, a
University of Sydney Merit Award and a Denison Merit Award.

\References{
\rfr Appourchaux, T. 2008, Astronomische Nachrichten, 329, 485
\rfr Bedding, T. R., Butler, R. P., Carrier, F., et al. 2006, ApJ, 647, 558
\rfr Bedding, T. R., Kjeldsen, H., Arentoft, T., et al. 2008, ApJ, 663, 1315
\rfr Benomar, O. 2008, Communications in Asteroseismology, 157, 98
\rfr Benomar, O., Appourchaux,R., \& Baudin, F. 2009, A\&A, 506, 15
\rfr Bretthorst, G. L. 1988, Lecture Notes in Statistics, Vol. 48, Bayesian Spectrum Analysis 
and Parameter Estimation (Springer-Verlag, New York)
\rfr Brewer, B. J., Bedding, T. R., Kjeldsen, H., \& Stello, D. 2007, ApJ, 654, 551
\rfr Brewer, B. J., \& Stello, D. 2009, MNRAS, 395, 2226
\rfr Carrier, F., \& Bourban, G. 2003, A\&A, 406, L23
\rfr Gaulme, P., Appourchaux, T., \& Boumier, P. 2009, A\&A, 506, 7
\rfr Gregory, P. C. 2005, Bayesian Logical Data Analysis for the Physical Sciences: A Comparative Approach
with `Mathematica' Support (Cambridge University Press)
\rfr Kjeldsen, H., \& Bedding, T. R. 1995, A\&A, 293, 87
\rfr Kjeldsen, H., Bedding, T. R., Butler, R. P., et al. 2005 ApJ, 635, 1281
\rfr Neal, R. M. 1993, Probabilistic Inference Using Markov Chain Monte Carlo Methods, Technical Report
CRG-TR-93-1, Dept. of Computer Science, University of Toronto, available at 
\url{http://www.cs.toronto.edu/~radford}
\rfr Roberts, D. H., Lehar, J., \& Dreher, J. W. 1987, AJ, 93, 968
\rfr Stello, D., Kjeldsen, H., Bedding, T.R., \& Buzasi, D. 2006, A\&A, 448, 709
\rfr Tassoul, M. 1980, ApJS, 43, 469
\rfr Tassoul, M. 1990, ApJ, 358, 313
}

\end{document}